\documentclass[12pt]{iopart} 
\usepackage{epsfig}
\usepackage{array}
\usepackage{psfrag}
\usepackage{amssymb}

\begin{document}
\title{Dynamical scaling for probe particles in a driven fluid}
\author{A R\'akos$^1$, E Levine$^2$, D Mukamel$^1$ and
G M Sch\"{u}tz$^3$} 
\address{$^1$ Department of Physics
of Complex Systems, Weizmann Institute of Science, Rehovot, Israel
76100.}
\address{$^2$ Department of Physics and Center for Theoretical
Biological Physics, University of California at San Diego,
California 92093-0319} 
\address{$^3$ Institut f\"{u}r
Festk\"orperforschung, Forschungszentrum J\"{u}lich, 52425
J\"{u}lich, Germany.}
\eads{\mailto{attila.rakos@weizmann.ac.il}, \mailto{levine@ctbp.ucsd.edu}, \mailto{david.mukamel@weizmann.ac.il} and \mailto{g.schuetz@fz-juelich.de}}


\begin{abstract}
We investigate two distinct universality classes for probe
particles that move stochastically in a one-dimensional driven
system. If the random force that drives the probe particles is
fully generated by the current fluctuations of the driven fluid,
such as when the probe particles are embedded in a ring,  they
inherit the dynamical exponent of the fluid, which generically is
$z=3/2$. On the other hand, if the random force has a part that is
temporally uncorrelated, the resulting motion can be described by
a dynamical exponent $z=2$ as considered in previous work.
\end{abstract}

\begin{indented}
\item[] {\bf Keywords:\/} driven diffusive systems (theory), stochastic particle dynamics (theory)
\end{indented}

\section{Introduction}
\label{intro}

The  effective interaction between probe particles in a fluid has
been a subject of theoretical and experimental studies for some
time. If a liquid is far from thermal equilibrium then the
interactions involving probe particles cannot be described by
classical thermodynamic approaches such as the potential
distribution theorem \cite{Wido63}. Even the large scale random
motion of a single particle cannot naively be assumed to be Brownian
motion as temporal correlations between successive collisions of
the probe particle with the fluid particles cannot safely be
ignored. At present too little is known in general terms about random
motion in non-equilibrium fluids.

Some progress may be achieved by considering stylized lattice gas
models of fluids. Such models sometimes are amenable to analytical
treatment and may thus give some insight into universal features
of the stochastic motion and the effective fluid-mediated
interaction of probe particles. A paradigmatic microscopic lattice
gas model is the asymmetric simple exclusion process (ASEP)
\cite{Ligg99,Schu01}, the large-scale dynamics of which are
governed by the Burgers equation \cite{Burg74}. In particular, in
one dimension the motion of so-called second class particles is
well understood. Such a probe particle moves with an average speed
given by the collective velocity $v_c = \partial j/\partial \rho$
where $j$ is the steady state current of the lattice gas with
density $\rho$. The fluctuations of the position around its mean
are superdiffusive with the dynamical exponent $z=3/2$ of the
universality class of the noisy Kardar-Parisi-Zhang (KPZ) equation
\cite{KPZ}. This behaviour is intricately linked to the nature of
the current fluctuations in the KPZ universality class. Moreover,
the second class particle is attracted to regions of large density
gradient (on molecular scale), thus serving as a microscopic
marker of shocks \cite{Shock}.

When a large system contains two probe particles, the local
density gradient in the density of the fluid particles induces an
effective long-range attractive interaction between the probes.
The steady-state distribution of the distance $x$ between the two
second class particles was found to decay as $x^{-3/2}$ for large
$x$ \cite{Derrida93}. In \cite{Kafr02} it was suggested that
this algebraic behaviour generalizes to $x^{-b}$ for other driven
diffusive systems with a non-universal exponent $b$. This leads to
phase separation for $b>2$. For the {\it dynamics} of two probe
particles, initially introduced into a pure system at a finite
distance, this implies an increase of the mean distance ($b<2$) or
the formation of bound state with finite mean distance ($b>2$). In
recent work \cite{Levi05} an attempt was made to determine the
stochastic motion of two probe particles by a Fokker-Planck
approach. This approach implies dynamical scaling with $z=2$ and
yields a differential equation for the scaling form of the
dynamical probability distribution for the distance of the probe
particles. In particular, one obtains the growth law for the mean
distance $\langle x(t)\rangle \propto t^\nu$ and the growth exponent
$\nu$ as a function of $b$.

Here we argue that the analysis of \cite{Levi05} applies under
certain conditions to the dynamics of probe particles which has a
random part that is {\it uncorrelated} in time. A simple model for
which such dynamics applies is given below. On the other hand, for
probe particles driven by stochastic forces originating solely
from the dynamics of the surrounding fluid particles, as in the
case of probe particles embedded in a ring, the scaling behaviour
is determined by the dynamical exponent of the fluid. Namely, one
has $z=3/2$ for generic driven lattice gases rather than the
non-generic Edwards-Wilkinson exponent $z=2$. This picture is
theoretically founded below in a more detailed consideration of
the statistical properties of the underlying current fluctuations
and supported by dynamical Monte-Carlo simulations.

The paper is organized as follows. In section~\ref{model} we define a
generalization of the ASEP with probe particles, originally
introduced in \cite{Kafr03}. Moreover we introduce a variant of
the model, with the same transitions, but driven by a different
source of noise. In section~\ref{KPZ} we use dynamical scaling to determine
the growth exponent $\nu$ for the driven fluid with $z=3/2$. In
section~\ref{segment} we analyze the dynamical behaviour of the modified model
which is governed by the $z=2$ dynamical exponent. In section~\ref{conc} we
reconsider the results of \cite{Levi05} and present our
conclusions.

\section{Models for driven fluids}
\label{model}

Following \cite{Kafr03} we define our model on a one dimensional (1d) lattice. A
hard core repulsion of particles is incorporated by an exclusion
interaction such that each lattice site can be occupied by either
a positive particle ($+$), a negative particle ($-$), or it may be
occupied by a probe particle ($0$). The continuous time stochastic
dynamics of the model is defined by nearest neighbor jumps where
the particles are exchanged with rates
\begin{eqnarray}
\label{eq:model}
+ - &\to - + \qquad &\mbox{with rate}~1+\Delta V \cr
+\ 0 &\to \, 0 + \qquad &\mbox{with rate}~1 \cr
\,0 - &\to -\ 0 \qquad &\mbox{with rate}~1\;.
\end{eqnarray}
Here $\Delta V$ is the energy of the initial configuration minus that of the final one, defined by the nearest neighbor Ising interaction
potential
\begin{equation} V = -
\frac{\epsilon}{4} \sum_i s_i s_{i+1}\;,
\end{equation}
where $s_i=0, \pm1$ according to the occupation of site $i$, and
$-1 \leq \epsilon \leq 1$. We consider the range $-0.8<\epsilon<1$
of the coupling constant. We mention that in the case where $\epsilon<-0.8$ (not studied here) the current-density relation of a domain of particles exhibits two degenerate maxima. This could lead to a phase separation of the two maximal current phases within the domains, which makes the analysis more involved. For a discussion of this case we refer to reference~\cite{Godr05}.

This model reduces to the KLS model \cite{KLS,Popk99} in the case
where no probe particles are present. Then the steady state has an
Ising measure \cite{KLS}. This is also expected to be the local
steady-state measure away from any probe when the density of
probes is zero. As in most studies of this model we consider equal
densities of positive and negative particles. In Monte Carlo
simulations to be discussed below, one considers $N$ sites with
periodic boundary conditions and random sequential update
dynamics.

In the absence of nearest neighbor interaction, $\epsilon=0$, the
dynamics defined above reduces to that of the TASEP with
second-class particles which are the probes. The steady state of
this system is fully known \cite{Derrida93}. The statistical
weight of all configurations with a given segment length $n$ between the two probe
particles is proportional to $Z_n$, the partition function of the
TASEP on an open chain of length $n$ in the maximal current phase
\cite{TASEP}. This observation has been used to show that the
probability to find the two probes at a distance $n$ from each
other decays for large $n$ as $n^{-3/2}$ \cite{Derrida93}. In
addition, this result can be used to estimate the currents of
particles which go in and out of the segment trapped between the
two probe particles. One finds that the outgoing current of $+$
($-$) particles through the right (left) probe takes the form of
the steady state current of the TASEP. To leading order in $1/n$
it is given by $j^{\rm
out}_n=\frac{1}{4}\left(1+\frac{3/2}n\right)$. The opposing
currents, namely that of $-$ ($+$) particles incoming through the
right (left) probe, take the form $j^{\rm
in}=\frac{1}{4}\left(1+\frac{3/2}{N-n-2}\right)$. In the limit we
are concerned with, namely $N \gg n$ and $N \to \infty$, this
current is well approximated by $j^{\rm in}=\frac{1}{4}$.

Following \cite{Kafr03} one expects similar behaviour for $\epsilon
\neq 0$. Namely, that the current of $+$ ($-$) particles bypassing
the right (left) probe takes the same form as that of the current
in an open segment of the same length, governed by the same
dynamics. This current is given by (again to leading order in
$1/n$)
\begin{equation}
\label{jout} j^{\rm
out}_n=j_0(\epsilon)\left(1+b(\epsilon)/n\right),
\end{equation}
where
\begin{equation}
\label{eq:j} j_0(\epsilon) = \frac{\upsilon +
\epsilon}{\upsilon^{3}} \;,\quad b(\epsilon)=\frac{3}{2}\;
\frac{(2+\epsilon)\upsilon+2\epsilon}{2(\upsilon+\epsilon)}\;,
\end{equation}
and $ \upsilon=\sqrt{\frac{1+\epsilon}{1-\epsilon}}+1$. For the
relevant values of $\epsilon$ one has $0\leq b\leq9/4$. Also,
similar to the $\epsilon=0$ case, the incoming current is given in
the large $N$ limit by
\begin{equation}\label{jin}
j^{\rm in}=j_0(\epsilon).
\end{equation}
Equation (\ref{jout}) has been derived in \cite{Krug90}, where $b$ is
proportional to the coefficient of the non-linear term in the KPZ
equation.

The dynamics of two probe particles embedded in an infinitely long
ring may therefore be rephrased as follows: The left probe
particle jumps to the left (thus increasing the size of the
cluster between the two probes by one lattice unit) after a random
time determined by the current of the (infinite) environment. The
mean of that random time is $\tau^{\rm in}=1/j^{\rm in}$. It jumps
to the right (thus decreasing the cluster size by one lattice
unit) after a random time with mean $\tau^{\rm out}(n)=1/j^{\rm
in}(n)$ which depends on the size of the cluster. The motion of
the right probe particles can be described similarly. Therefore
the distance $n$ between the two probes is a random process which
is increased by one unit by the incoming current. This happens
after a random time with mean $\tau^+=\tau^{\rm in}/2$,
independent of the cluster size. The distance is decreased by one
unit after a cluster size dependent random time with mean
$\tau^-(n)=\tau^{\rm out}(n)/2$.

We stress that the incoming and outgoing currents are correlated
in time and hence do {\it not} generate a memoryless Markovian
random motion of the probe particles. As discussed below the
dynamics of the probe particles is controlled in this case by the
dynamical exponent $z=3/2$ of the KPZ universality class.

An interesting modification of the model is given by the following
stochastic dynamics: The motion of the probe particles that lead
to an increase of the cluster is taken to be a Poisson process
with constant mean waiting time $\tau^{\rm in}$. The motion that
leads to a decrease of the cluster size is unchanged in the sense
that it is generated by the jump events inside the cluster. Thus
while the combined stochastic motion of the probe particles is
also non-Markovian, the noise has a contribution that is
uncorrelated in time. Below we shall refer to this dynamics as
mixed.

This mixed dynamics can be described within a similar framework of
the model defined above. Consider a fluid segment of fluctuating
length composed of $+/-$ particles denoted by $\mathbf{A}$ embedded in an environment of 
``probe particles'' $0$. The bulk dynamics of the particles within
the segment is defined as before by the KLS hopping rules. The
left boundary dynamics of a cluster $\mathbf{A}$ with length $n$ is defined
as follows:
\begin{eqnarray}
\label{eq:left_boundary}
0\ 0\ \mathbf{A}\; &\to\; 0 + \mathbf{A} \qquad &\mbox{with rate}~j_0(\epsilon) \cr
0 - \mathbf{A}\; &\to\; 0\ 0\ \mathbf{A} \qquad &\mbox{with rate}~1.
\end{eqnarray}
Similarly at the right boundary one takes
\begin{eqnarray}
\label{eq:right_boundary}
\mathbf{A}\ 0\ 0\; &\to\; \mathbf{A} - 0 \qquad &\mbox{with rate}~j_0(\epsilon) \cr
\mathbf{A} + 0\; &\to\; \mathbf{A}\ 0\ 0 \qquad &\mbox{with rate}~1.
\end{eqnarray}
Note that a fluid segment of length $n=1$ can only increase in
size. This model is considered in more detail in section~\ref{segment}.

\section{Probe particles in a KPZ fluid}
\label{KPZ}

In the steady state the large distance distribution of the two
probes decays algebraically $\propto 1/n^b$ with the non-universal
exponent $b$ given in (\ref{eq:j}). For $b<2$ this leads to an
infinite mean distance as time tends to infinity. In order to
calculate the temporal evolution of the asymptotic mean distance
we make a scaling ansatz
\begin{equation}
\label{scaling} P(x,t) = x^{-b}t^{-\beta}f(x/t^{1/z})
\end{equation}
for the large scale probability density of the distance $x$
between the probes at time $t$. In this scaling picture the distance is considered to be continuous and correspondingly the notation is changed from $n$ to $x$. The scaling function $f(u)$ satisfies the boundary condition $f(0)= \mbox{finite}$,  and
the normalization condition $\int_\theta^\infty P(x)\,\rmd x = 1$.
Here a cutoff $\theta$ is introduced to prevent divergence at
$x\to 0$ for $b\geq1$. This generalizes the scaling ansatz of 
reference~\cite{Levi05}.

In the range $b>1$ the normalization condition yields $\beta=0$
while for $b<1$ one finds $\beta=(1-b)/z$. From this we calculate
the mean distance between the particles, $ \langle x(t)\rangle =
{\int_\theta^\infty x P(x,t)\,\rmd x}$. Inspecting the large $t$
behaviour of this integral one finds
\begin{equation}
\label{eq:nu} \langle x(t)\rangle \sim
\cases{
t^{1/z}& $b<1$ \\
t^{1/z}/\log(t)& $b=1$ \\
t^{(2-b)/z}& $1<b<2$ \\
\log(t)& $b=2$ \\
A + t^{-(b-2)/z} & $b>2$}
\end{equation}
Here $A$ is a non-universal constant. Hence for $b<2$
the mean distance increases algebraically in time with exponent
$\nu = 1/z$ ($b<1$) or $\nu = (2-b)/z$ ($1<b<2$) respectively. At
$b=1,2$ there are logarithmic corrections. For $b>2$ the two
particles are bound to each other with a finite average distance while
for $1<b<2$ the particles are weakly bound.

Since the motion of the probe particle is driven by the
fluctuations of the current we argue that the scaling form of the
probability distribution for the distance between two probes
should be determined by the dynamical exponent of the fluid. This
leads to the identification of $z$ with the KPZ dynamical
exponent. Let us briefly elaborate on this point. Consider the
position of, say, the left probe particle. The rate of change of
its position is given by the difference between the incoming
currents into the segment confined by the probes and the outgoing
currents from this segment at the left end. The incoming current
originates from a very large KPZ fluid and hence the variance of
this quantity grows as $t^{4/3}$ \cite{Henk}. The outgoing current
originates from the segment of finite length, and is highly
correlated with the incoming currents. In the following we show
that the distribution of the position of the probe particle is
determined by the KPZ exponent of the incoming current, $z=3/2$.
To this end we study numerically the fluctuations of the incoming
and the outgoing currents at, say, the left end separately. We
define $X_{L+}(t)$ as the number of steps of the left probe which
increase the distance between the probes up to time $t$. These are
the moves to the left. Similarly we define $X_{L-}(t)$ as the
number of length decreasing steps up to time $t$, i.e. the time
integrated number of steps to the right. Analogously we define the
quantities $X_{R\pm}(t)$ for the right probe.

\begin{figure}
\psfrag{XLABEL}[cc][cc]{$t$}
\psfrag{LABEL1}[cl][cl]{$\langle X_{L+}^2 \rangle_c$}
\psfrag{LABEL2}[cl][cl]{$\langle X_{L-}^2 \rangle_c$} 
\psfrag{LABEL3}[cl][cl]{$\langle X_{L+}X_{R-} \rangle_c$} 
\psfrag{LABEL4}[cl][cl]{$-\langle X_{L+}X_{R+} \rangle_c$}
\psfrag{LABEL5}[cl][cl]{$-\langle X_{L+}X_{L-} \rangle_c$}
\psfrag{LABEL6}[cl][cl]{$\sim t^{4/3}$}
\psfrag{LABEL7}[cl][cl]{$-\langle X_{L-}X_{R-} \rangle_c$}
\psfrag{LABELA}[cl][cl]{(a)}
\psfrag{LABELB}[cl][cl]{(b)}
\begin{center}
\epsfig{file=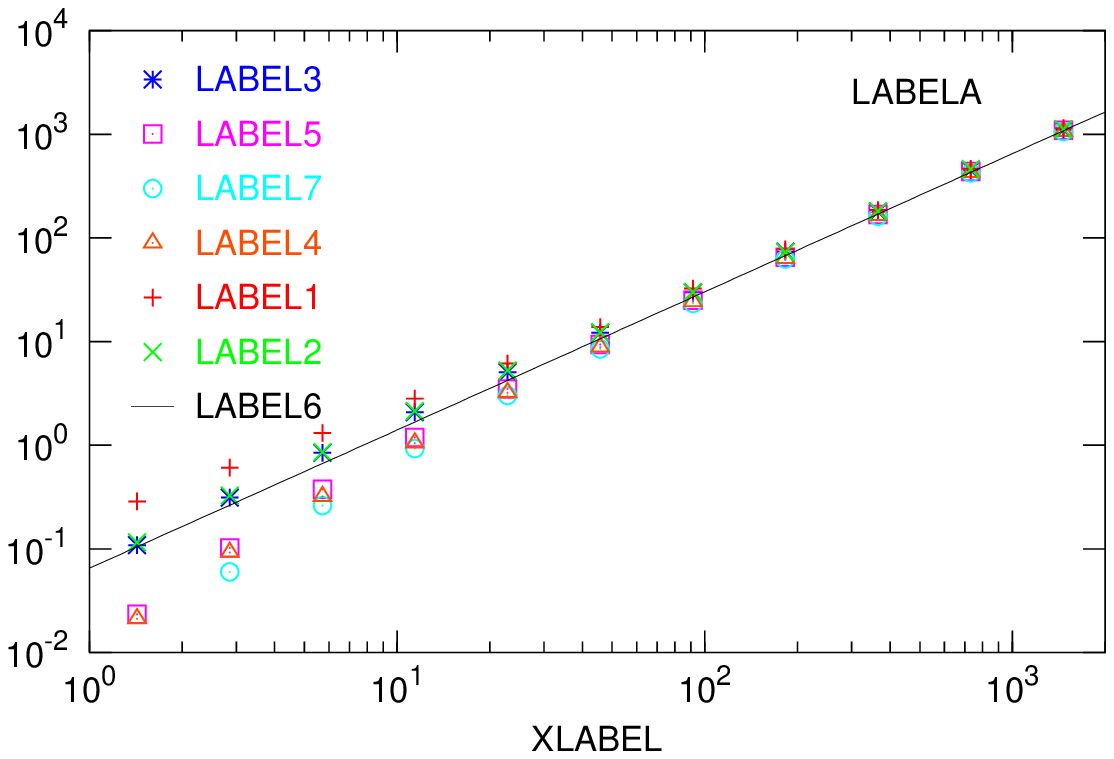, angle=270}
\epsfig{file=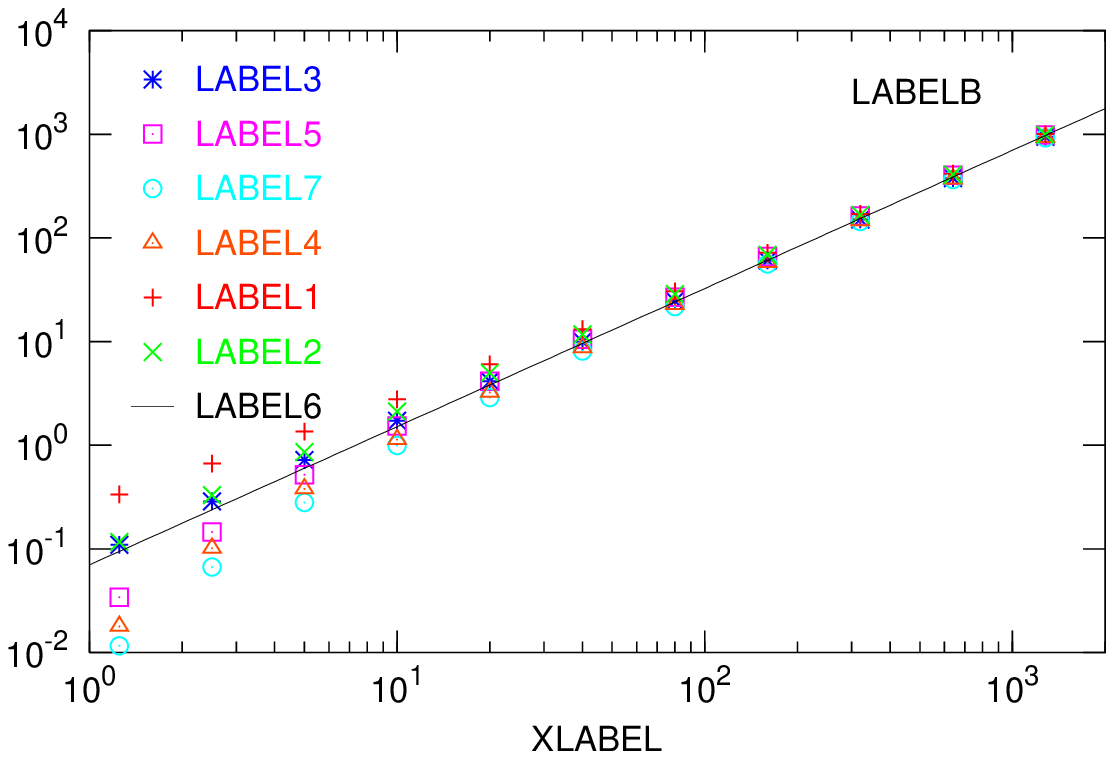, angle=270} 
\end{center}
\caption{\label{fig:xx}
Measured variances and covariances of the quantities
$X_{L+}$, $X_{R+}$, $X_{L-}$, and $X_{R-}$ for (a) $\epsilon=0.4$
$(b=1.75)$ and (b) $\epsilon=-0.6$ $(b=0.75)$. $\langle AB \rangle_c$ denotes $\langle AB \rangle - \langle A\rangle \langle B\rangle$. Note that because
of the left-right symmetry some of the possible combinations are
equal to each other. The simulation parameters are $N=1600$, 10000
realizations and equilibration times (see below)
$t_\mathrm{equil}=14286$ (a), $t_\mathrm{equil}=12500$ (b).}
\end{figure}

We have calculated the variances and the covariances of the four
random variables $X_{L\pm}$ and $X_{R\pm}$ in simulations. We find
that all these quantities increase to leading order as
$t^{4/3}$ (see figure~\ref{fig:xx}). This demonstrates that the
dynamical exponent is $z=3/2$.

The scaling ansatz (\ref{scaling}) predicts for $1<b<3$ that the variance $\sigma^2$ of the distance $x(t)=X_{L+}(t) + X_{R+}(t) - X_{L-}(t) - X_{R-}(t)$ between the two probes increases with a lower power of $t$, namely, $\sigma^2 \propto t^{(3-b)/z}$. This requires the cancellation of the leading term contribution of $\sigma^2$, which is in agreement with the results of the numerical simulations presented in figure~\ref{fig:xx}(a), where the amplitudes of all the variances and covariances appear to be the same.  

\begin{figure}
\psfrag{XLABEL}[cc][cc]{$t$} 
\psfrag{YLABEL}[cc][cc]{$\langle x \rangle$} 
\psfrag{LABEL8}[cl][cl]{$\epsilon = 0.8 \quad b=1 $}
\psfrag{LABEL4}[cl][cl]{\makebox[1.7cm][l]{$\epsilon = 0.4$} \makebox[1.8cm][l]{$b=1.759$} $\nu=0.161$}
\psfrag{LABEL0}[cl][cl]{\makebox[1.7cm][l]{$\epsilon = 0$} \makebox[1.8cm][l]{$b=3/2$} $\nu=1/3$}
\psfrag{LABELm3}[cl][cl]{\makebox[1.7cm][l]{$\epsilon = -0.3$} \makebox[1.8cm][l]{$b=1.22$} $\nu=0.515$}
\psfrag{LABELA}[cl][cl]{}
\begin{center}
\epsfig{file=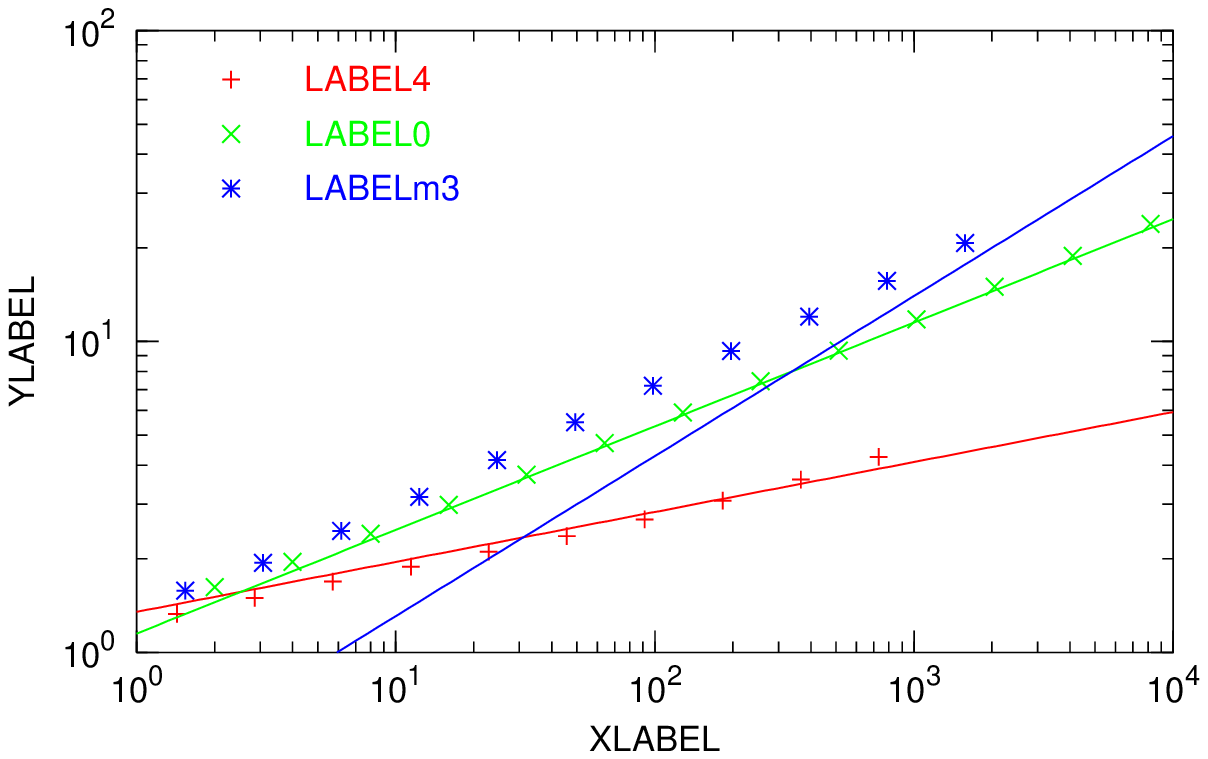, angle=270}
\end{center}
\caption{\label{fig:average_distance} Average distance of the two
probes as a function of $t$ for some values of $\epsilon$. Solid
lines indicate the predicted behaviour for large $t$. The data are
shown on a log-log scale. The simulation parameters are $\epsilon
= 0.4$: $N=1000$, $t_\mathrm{equil}=14286$, and 64000 realizations;
$\epsilon = 0$: $N=7000$, $t_\mathrm{equil}=100000$, and 64000
realizations; $\epsilon = -0.3$: $N=1600$, $t_\mathrm{equil}=15385$,
and 24160 realizations.}
\end{figure}

The mean distance is predicted to grow as
$t^{(2-b)/z}$ for $1<b<2$. This is confirmed by Monte Carlo data presented in
figure~\ref{fig:average_distance} for some values of $b$ in this
range. At $b=2$ $(\epsilon=0.8)$ the mean distance is predicted to
grow logarithmically in time. Our numerical simulations for this
case show that the mean distance remains very small (of the order
of a few lattice constants) in the time interval which was
studied, and it grows slowly in time.

To complete our analysis we measured the cumulative distribution
function
\begin{equation}
\tilde P(x,t) = \mathrm{Prob}(\mbox{distance between the two
probes } > x \mbox{ at time } t).
\end{equation}
For this quantity, the scaling ansatz (\ref{scaling}) yields
\begin{equation}
\label{scalingform} \tilde P(x,t) = t^{(1-b)/z-\beta}
F(xt^{-1/z}),
\end{equation}
where $F(y) = \int_y^{\infty} \eta^{-b} f(\eta) d\eta$.
Consequently, in the case of $b>1$, $F(y)$ is expected to behave
as $y^{-b+1}$ for small $y$, whereas $b<1$ implies $F(0)=1$.

\begin{figure}
\psfrag{XLABEL}[cc][cc]{$y = x t^{-2/3}$}
\psfrag{TITLEa}[cc][cc]{(a) $\epsilon=0.8$,  $b=2$}
\psfrag{YLABELa}[cc][cc]{$\tilde P(x,t) t^{2/3}$}
\psfrag{LABEL1a}{$\sim y^{-1}$}
\psfrag{TITLEb}[cc][cc]{(b) $\epsilon=0.4$, $b=1.759$}
\psfrag{YLABELb}[cc][cc]{$\tilde P(x,t) t^{0.506}$}
\psfrag{LABEL1b}{$\sim y^{-0.759}$}
\psfrag{TITLEc}[cc][cc]{(c) $\epsilon=0$, $b=1.5$}
\psfrag{YLABELc}[cc][cc]{$\tilde P(x,t) t^{1/3}$}
\psfrag{TITLEd}[cc][cc]{(d) $\epsilon=-0.3$, $b=1.228$}
\psfrag{YLABELd}[cc][cc]{$\tilde P(x,t) t^{0.152}$}
\psfrag{LABEL1d}{$\sim y^{-0.228}$}
\psfrag{LABEL1c}{$\sim y^{-1/2}$}
\begin{center}
\epsfig{file=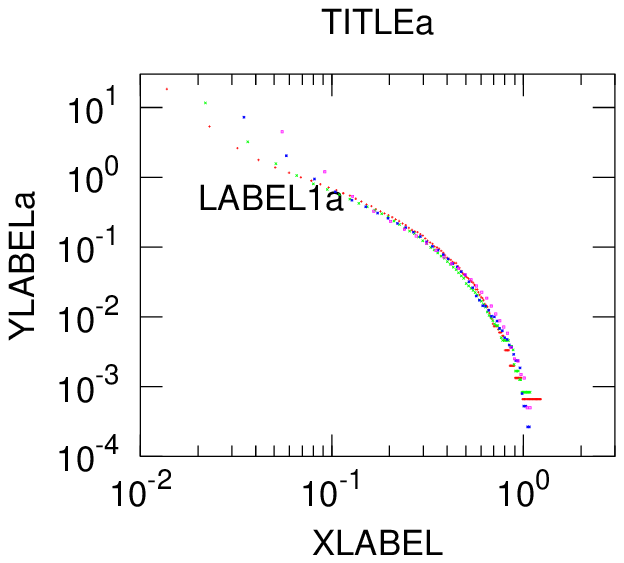, angle=270}~\epsfig{file=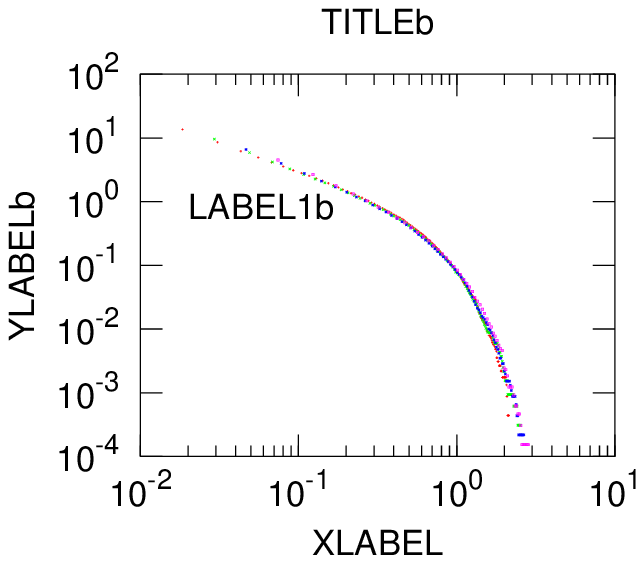, angle=270}

\vspace{0.5cm}
\epsfig{file=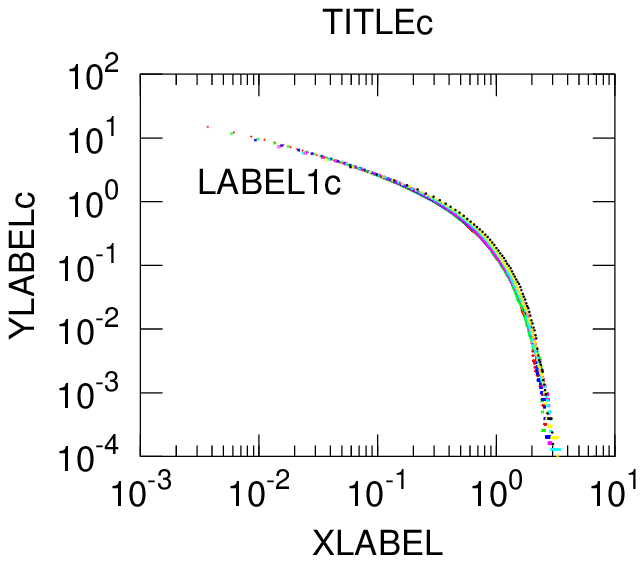, angle=270}~\epsfig{file=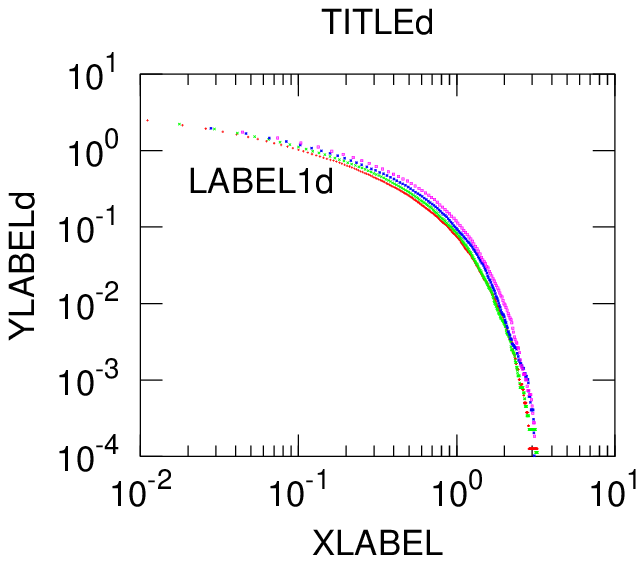, angle=270}
\end{center}
\caption{\label{fig:data_collapse} Simulation results for the KPZ
fluid with two probes. The cumulative distribution function
$\tilde P(x,t)$ of the distance between the two probes is measured
and the quantity $\tilde P(x,t) t^{\frac{2}{3}(b-1)}$ is plotted
against $xt^{-2/3}$ for some values of the parameter
$\epsilon$. The simulations were carried out using an efficient
multispin coding algorithm. The simulation parameters are: (a)
$N=1600$, $t_\mathrm{equil}=11111$, $t=142, 284, 569, 1138$ and
163712 realizations; (b) $N=1000$, $t_\mathrm{equil}=14286$, $t=91,
183, 366, 731$ and 64000 realizations = 64000; (c) $N=7000$,
$t_\mathrm{equil}=100000$, $t=128, 256, 1024, 2048, 4096, 8192$ and
64000 realizations; (d) $N=1600$, $t_\mathrm{equil}=15385$, $t=197,
394, 788, 1575$ and 24160 realizations.}
\end{figure}

The results for the scaling ansatz for the distribution function
are shown in figure~\ref{fig:data_collapse}, where the quantity
$\tilde P(x,t) t^{\frac{2}{3}(b-1)}$ is plotted against
$xt^{-2/3}$ for some values of $b>1$. The data collapse shows that
in this regime the scaling ansatz is correct with $z=3/2$ and
$\beta=0$. On the other hand for $0<b<1$ no good data collapse is
found. This may be understood by noting that at $b=0$ the leading
quadratic nonlinearity in the KPZ equation vanishes and one is
left with a quartic nonlinearity. According to standard wisdom,
power counting shows that such a nonlinear term is irrelevant in
the renormalization group sense. One expects a dynamical
Edwards-Wilkinson exponent $z=2$, even though the fluid is driven.
This results in strong crossover effects for small $b$, making the
scaling form (\ref{scalingform}) with $z=3/2$ invalid.
Measurements of the fluctuations of the individual probe particles
(not shown in this paper) are found to be consistent with $z=2$.

A subtle point of the simulation is the proper choice of the
initial condition. We simulated the process on a ring of $N$
(even) sites with equal ($N/2-1$) number of $+$ and $-$ particles
and two probes. Initially the two probes were placed on
neighbouring sites and any configuration of the remaining $N-2$
particles was chosen with equal probability.
The process (\ref{eq:model}) was then simulated up to some time
$t_\mathrm{equil}$ with the modification that the two probes are
constrained to remain nearest neighbours, i.e., these particles
hop together as if they occupied only a single site. During this
``equilibration'' the fluid approaches its stationary structure in
the vicinity of the probes up to a distance of the order
$t_\mathrm{equil}^{2/3}$. At this point, defined as $t=0$, the
restriction for the relative position of the probes is released
and the distance between the two probes is measured. In our
analysis we consider $t\ll t_\mathrm{equil}$ so that the probe
particles always move within an equilibrated region.

Clearly, there are also other relevant choices for the initial condition, which describe different physical scenarios. For example one may consider an initial state obtained by letting the two probes move freely and $t=0$ is defined as the time they first meet after some equilibration time. This case is not studied here. We believe that such a change in the initial condition will not change the scaling exponent, although it may result in a different scaling function. In order to test this statement we carried out Monte Carlo simulations of the $\epsilon=0$ case with equal number of $+$ and $-$ particles. The initial configuration was picked randomly with uniform probability conditioned on having the two probes on neighbouring sites. This initial state is equivalent to the above studied one with $t_\mathrm{equil}=0$. Results of the simulations (not presented here) yield the same scaling exponent $z=3/2$ as before, but the scaling function is different from the one in figure~\ref{fig:data_collapse}.c.

For more than two probes our dynamical scaling approach suggests the following scaling form for the joint distribution:
\begin{equation}
\label{scaling_several}
P(x_1, x_2, \cdots, x_M) = x_1^b x_2^b \cdots x_M^b t^{-M\beta} P(x_1 t^{-\frac1z}, x_2 t^{-\frac1z}, \cdots, x_M t^{-\frac1z})
\end{equation}
for $M+1$ probes with $z=3/2$. Here the formation of bound states for $b>2$ leads to a condensation transition for a finite density of probes. This phase transition was discussed in \cite{Kafr03}.

\section{Probe particles with mixed dynamics}
\label{segment}

In this section we consider the model of a cluster with mixed
dynamics and study its temporal evolution. In particular we are
interested in the growth law governing the length of the cluster.
As is evident from the definition of the model, two processes
control the dynamics of a cluster of length $x$: at both ends there is (a) a growth
process with rate $j_0(\epsilon)$, and (b) a length
decreasing process with rate $1$, see (\ref{eq:left_boundary}-\ref{eq:right_boundary}). To analyze the dynamics of the
cluster length we introduce processes for the individual motion of
the probes as defined above. The growth process at both ends is a
Poisson process and hence uncorrelated in time. Thus $\langle X_{R+}(t)\rangle
= \langle X_{R+}^2(t)\rangle -\langle X_{R+}(t)\rangle^2=j_0 t$ and similarly for
$X_{L+}$.

 The length decreasing processes,
$X_{R-}(t)$ and $X_{L-}(t)$  are correlated in time through the
dynamics of the particles within the cluster. Therefore the
process $x(t)$ is also non trivially correlated in time. The mean
rate of change of the length decreasing process,
${\rmd/\rmd t}{\langle X_{R-}(t)+X_{L-}(t)\rangle}=2j_x^{\rm out}$, depends on $x$
(\ref{jout}).

\begin{figure}
\begin{center}
\psfrag{L+L+}[cl][cl]{$\langle X_{L+}^2 \rangle_c$}
\psfrag{L-L-}[cl][cl]{$\langle X_{L-}^2 \rangle_c$} 
\psfrag{R+L-}[cl][cl]{$\langle X_{L+}X_{R-} \rangle_c$} 
\psfrag{R+L+}[cl][cl]{$\langle X_{L+}X_{R+} \rangle_c$}
\psfrag{-L+L-}[cl][cl]{$-\langle X_{L+}X_{L-} \rangle_c$}
\psfrag{j0t}[cl][cl]{$j_0 t$}
\psfrag{-L-R-}[cl][cl]{$-\langle X_{L-}X_{R-} \rangle_c$}
\psfrag{(a)}[cl][cl]{(a)}
\psfrag{(b)}[cl][cl]{(b)}
\epsfig{file=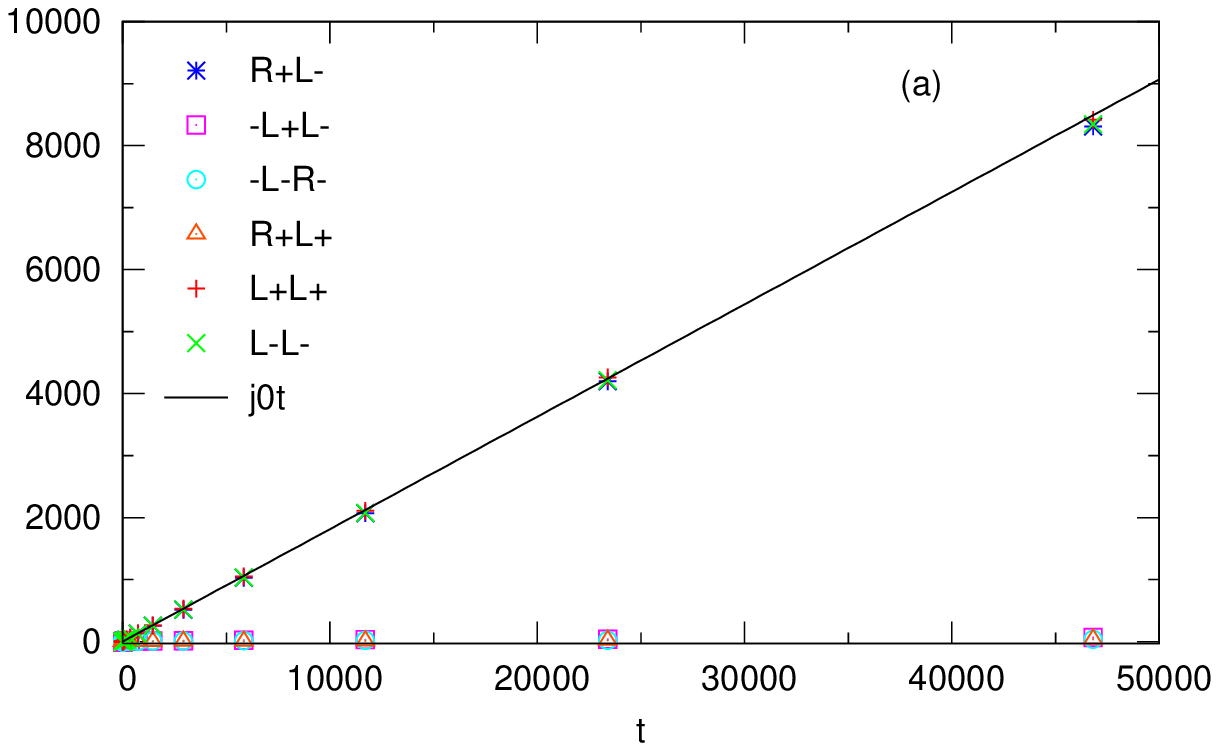,angle=270}
\epsfig{file=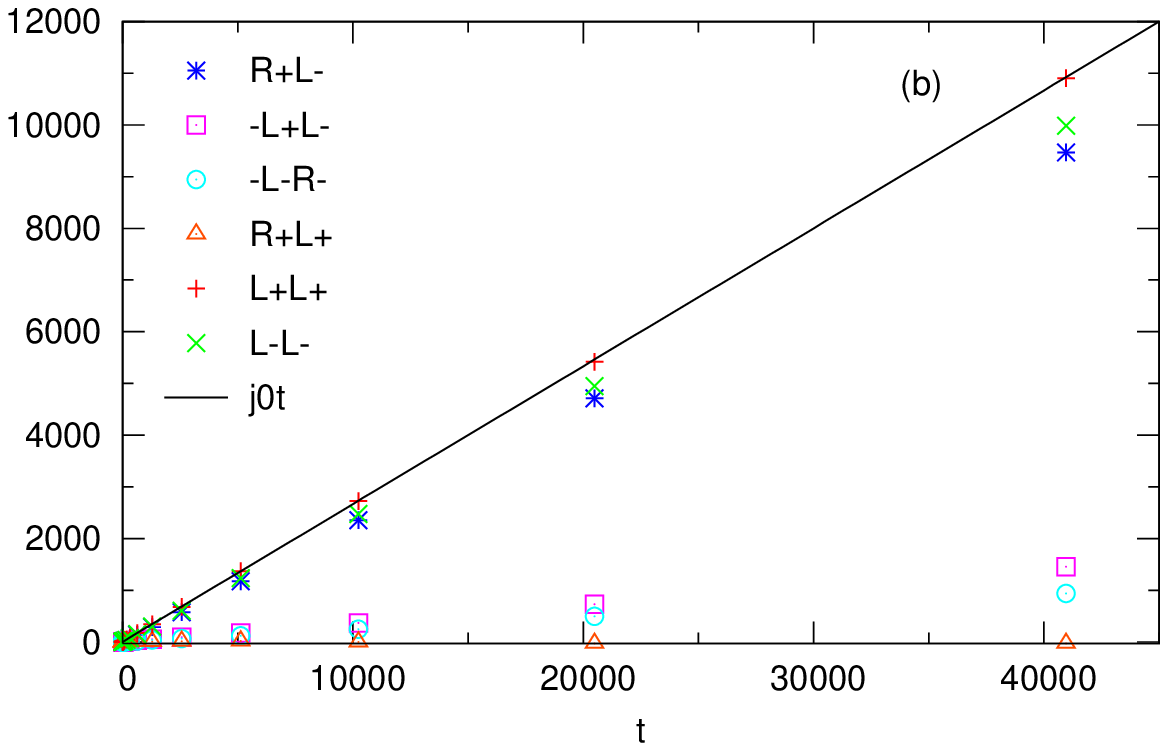,angle=270}
\end{center}
\caption{\label{fig:ZZZ} Measured variances and covariances of the quantities
$X_{L+}$, $X_{R+}$, $X_{L-}$, and $X_{R-}$ in the mixed model for (a) $\epsilon=0.4$
$(b=1.75)$ and (b) $\epsilon=-0.6$ $(b=0.75)$. $\langle AB \rangle_c$ denotes $\langle AB \rangle - \langle A\rangle \langle B\rangle$. Note that because
of the left-right symmetry some of the possible combinations are
equal to each other. The averages were taken over 10000 realizations.}
\end{figure}

\begin{figure}
\begin{center}
\epsfig{file=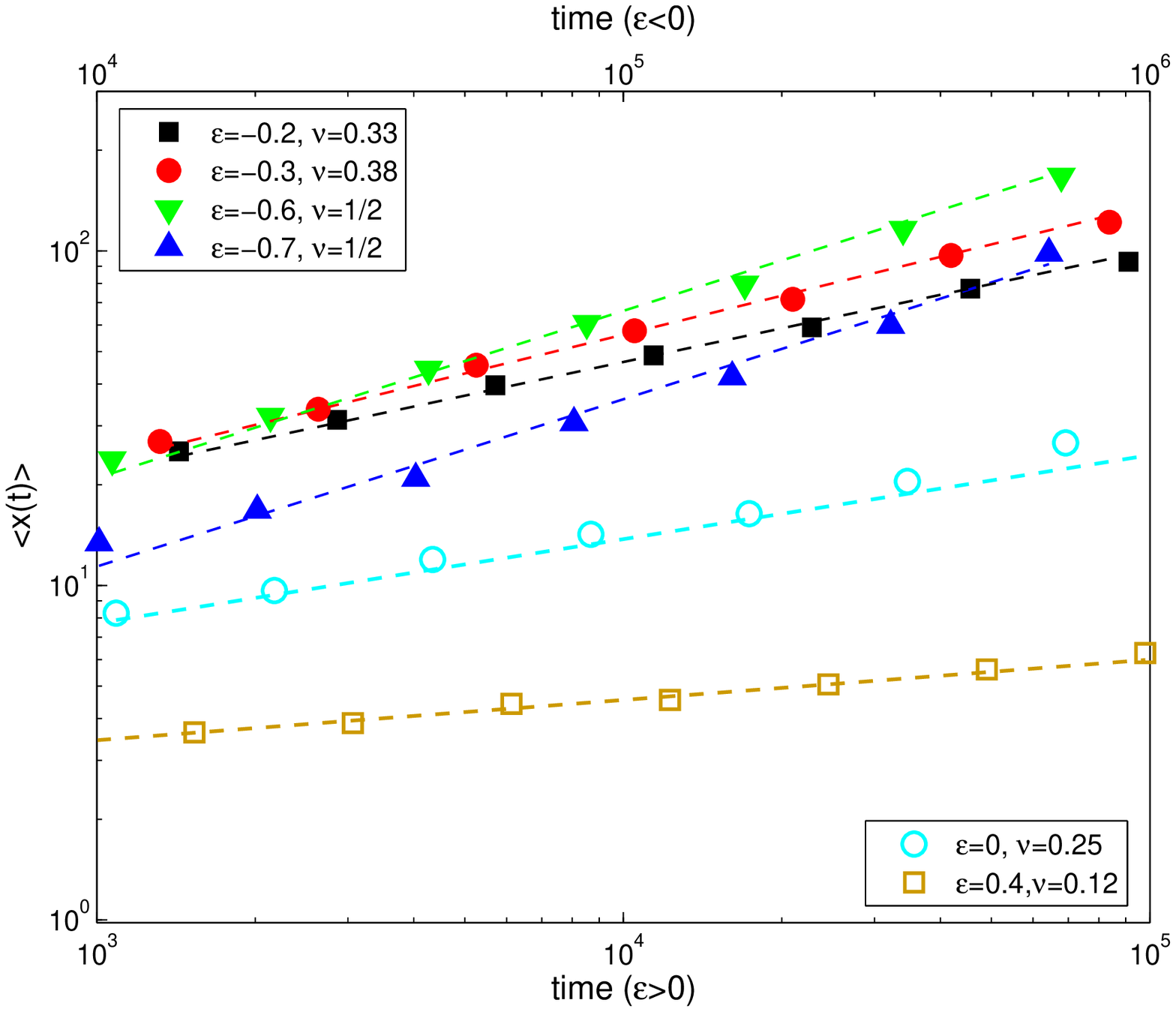, width=11truecm}
\end{center}
\caption{\label{fig:mixed-distance} Results of numerical
simulations for the average distance $\langle x(t)\rangle$ between the probes
for various values of $b$. Lines are drawn with a slope as
expected from equation~\ref{eq:nu}.}
\end{figure}

We start the analysis of $x(t)$ by demonstrating that the
dynamical exponent of this process is $z=2$. As before, we measure
the variances and covariances of $X_{R\pm}(t)$ and $X_{L\pm}(t)$.
We find that for all these quantities the asymptotic growth is linear in $t$. In combining these expressions to calculate $\sigma^2$ the leading order terms cancel for $b>1$ and the growth of $\sigma^2$ becomes sub-linear, just like in the case of the KPZ fluid. The fluctuations of
the motion of the probe particles are thus determined by the
fluctuations of the uncorrelated incoming current (see figure~\ref{fig:ZZZ}).
The temporal behaviour of the average distance between the probes
for various values of $\epsilon$ is given in
\fref{fig:mixed-distance}\footnote{Similar data for other values
of $b$ are given in reference~\cite{Levi05}. In that reference
the figure was erroneously referred to as corresponding to the KPZ
fluid rather than to the mixed dynamics.}. The measured value of the growth exponent $\nu$ obtained
from equation~(\ref{eq:nu}) is consistent with dynamical exponent
$z=2$.

\begin{figure}
\psfrag{XLABEL}[cc][cc]{$x t^{-1/2}$}
\psfrag{TITLEa}[cc][cc]{(a) $\epsilon=0$,  $b=1.5$}
\psfrag{YLABELa}[cc][cc]{$\tilde P(x,t) t^{1/4}$}
\psfrag{TITLEb}[cc][cc]{(b) $\epsilon=-0.4$, $b=1.104$}
\psfrag{YLABELb}[cc][cc]{$\tilde P(x,t) t^{0.0522}$}
\psfrag{TITLEc}[cc][cc]{(c) $\epsilon=-0.5$, $b=0.951$}
\psfrag{YLABELc}[cc][cc]{$\tilde P(x,t)$}
\psfrag{TITLEd}[cc][cc]{(d) $\epsilon=-0.6$, $b=0.75$}
\psfrag{YLABELd}[cc][cc]{$\tilde P(x,t)$}
\begin{center}
\epsfig{file=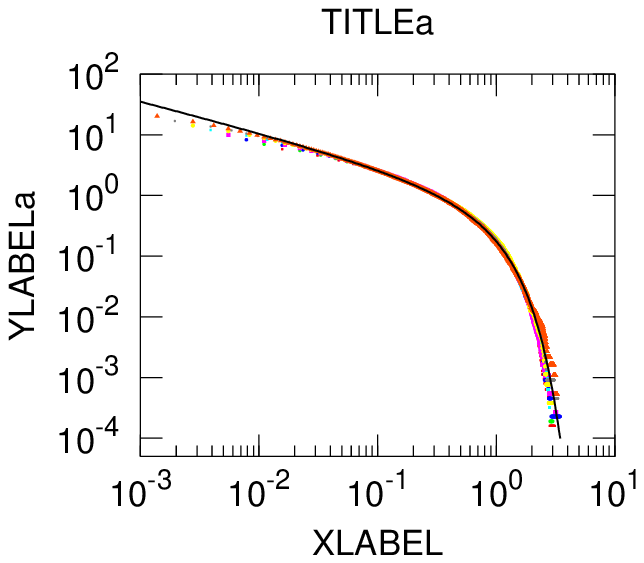, angle=270}~\epsfig{file=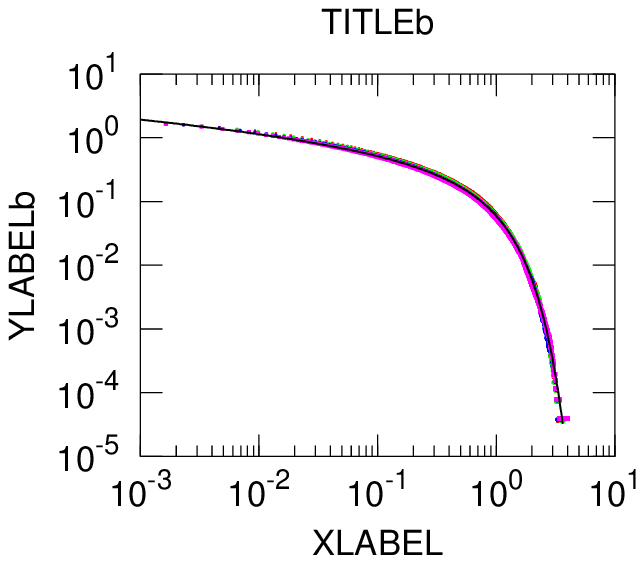, angle=270}

\vspace{0.5cm}
\epsfig{file=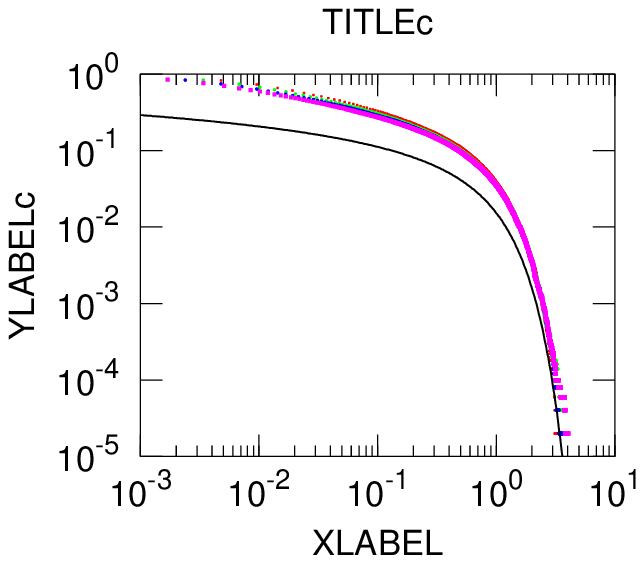, angle=270}~\epsfig{file=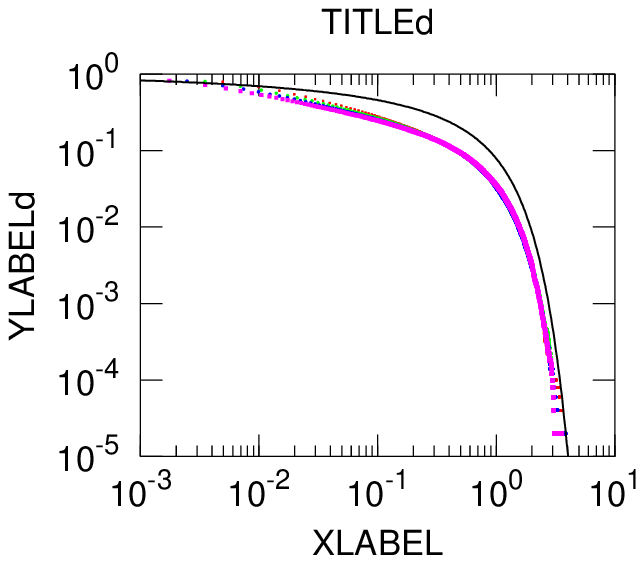, angle=270}
\end{center}
\caption{\label{fig:data_collapse_mixed} Simulation results for
the mixed model. The cumulative distribution function $\tilde
P(x,t)$ of the distance between the two probes is measured and the
quantity $\tilde P(x,t) t^{\frac{1}{2}(b-1)}$ (for $b>1$) and $\tilde P(x,t)$ (for $b<1$) is plotted against
$xt^{-1/2}$ for some values of the parameter $\epsilon$. The solid
line indicates the prediction of the Fokker-Planck equation which
is in good agreement with the data collapse for $b>1$. The
simulation parameters are: (a) $t=4096, 8192, 16384, 32768, 65536,
131072, 262144, 524288$; (b) $t=46811, 93622, 187246, 374491$; (c)
$t=43691, 87381, 174763, 349525$; (d) $t=40960, 81920, 163840,
327680$; and 50000 realizations for each case}
\end{figure}

In characterizing the dynamics of the two probes one may argue that they can be considered as two coupled random walkers. The driven fluid in between the probes generates an effective interaction, thus reducing the model to a two-particle problem. These particles hop away from each-other with a constant rate $j^{\rm in}$ of (\ref{jin}). On the other hand the rate at which they hop towards each-other, $j^{\rm out}_x$ , depends on the distance $x$ between them according to (\ref{jout}). This description  neglects the fact that the two-particle process is not memoryless and leads to a reduced master equation for the interparticle distance, which can be transformed into a Fokker-Planck equation in a continuum limit. This approach was taken in reference~\cite{Levi05} and the resulting scaling forms for the distance distribution were calculated. Here we compare this analytical result with simulations of the mixed model.
In the full range of $b<2$ we obtain good data collapse with a dynamical exponent $z=2$, which is consistent with the Fokker-Planck approach, see figure~\ref{fig:data_collapse_mixed}. However, For $b<1$ the scaling function does not seem to agree with the Fokker-Planck prediction of reference~\cite{Levi05}. We have no explanation
for this discrepancy.

An interesting generalization of the mixed model is obtained by allowing for probe particles between the fluctuating boundaries of the system. These extra probes have the same dynamics as the $0$ particles in the KPZ fluid (\ref{eq:model}) while the evolution at the boundaries is unchanged (\ref{eq:left_boundary}-\ref{eq:right_boundary}). Here the extra probes divide the system into several clusters. Numerical simulations (not presented here) show that the evolution of these clusters is described by the same dynamical exponent $z=2$ as for the case of no intermediate probes. We measured the quantities $X_{L+}, X_{R+}, X_{L-}, X_{R-}$ in a simulation for intermediate clusters and the result is very similar to those shown in figure \ref{fig:ZZZ}. Thus the uncorrelated noise acing on the boundaries dominate the entire system, including the inner probes which are not affected directly by it. These results suggest a scaling form similar to (\ref{scaling_several})
with $z=2$ for a system with $M$ clusters ($M-1$ inner probes). 

\section{Conclusions}
\label{conc}

In this paper the dynamics of two probe particles embedded in a
one dimensional driven fluid is studied. The non-equilibrium
dynamics of the fluid generates an effective interaction between
the probes. A simple model is introduced to study this dynamics.
Within the framework of this model, sufficiently strong attraction
between the fluid particles leads to a bound state of the probe
particles. For less pronounced attraction (or repulsion) a weakly
bound state is found in which the average distance between the
probes grows algebraically with time, with a non-universal growth
exponent. The probes become unbound for sufficiently strong
repulsion of the fluid particles, leading to dynamics
characterized by a universal growth exponent determined by the
dynamical exponent $z$ of the fluid.

In previous work by some of the authors of the present paper
\cite{Levi05} a similar analysis was performed using a Fokker-Planck 
approach, implying a dynamical exponent $z=2$. Here we
argue that this analysis does not apply to the KPZ fluid which was
envisaged in \cite{Levi05}. In order to clarify the validity of
that analysis, two cases which lead to different dynamical
exponents are considered here.

In the first case the probe particles are embedded in a KPZ like
fluid. Here the dynamical exponent which determines the motion of
the probes is predicted to be the KPZ exponent $z=3/2$. This is
verified by numerical studies.

In the second case the system is coupled to an external driving
force which is temporally uncorrelated. This leads to mixed
dynamics where both the temporally correlated noise generated by
the KPZ fluid and the uncorrelated external noise drive the motion
of the probes. Here we observe numerically that the dynamical
exponent is $z=2$. In this case the analysis of \cite{Levi05}
applies in the range $b>1$.

The different dynamical exponents of the two models can be traced
back to the fluctuations of the incoming and outgoing currents. We
observe that in both cases the exponent is determined by the
incoming current. This result is found to hold when the two models are generalized to include any finite number of probe particles. It would be interesting to understand how the
interplay of the correlated and the uncorrelated noise in the
mixed model generates $z=2$. Also the crossover behaviour for $b<1$
that leads to poor scaling collapse in the KPZ case is an
intriguing open problem.

\ack
We thank J. Krug, M. Barma and S. Chatterjee for helpful
discussions. Support of the Israel Science Foundation (ISF) and
the Minerva Einstein Center for Theoretical Physics is gratefully
acknowledged. Research of EL is supported in part by the NSF PFC-sponsored Center for Theoretical Biological Physics (Grants No.~PHY-0216576 and PHY-0225630). We thank the Newton Institute in Cambridge (UK) for
the kind hospitality during the programme "Principles of the
Dynamics of Non-Equilibrium Systems" where part of this work was
carried out.

\end{document}